\documentclass[12pt]{iopart}
\usepackage{graphicx}
\begin{document}

\title{Emergence of Global Preferential Attachment From Local Interaction}

\author{Menghui Li$^{1,2}$, Liang Gao$^{3}$, Ying Fan$^{1}$, Jinshan Wu$^{1,4*}$, Zengru Di$^{1}$}

\address{$^1$ Department of Systems Science, School of Management, Center for Complexity
Research, Beijing Normal University, Beijing 100875, P. R. China}
\address{$^2$ Temasek Laboratories, National University of Singapore, 117508,
Singapore \\
Beijing-Hong Kong-Singapore Joint Centre for Nonlinear
\& Complex Systems (Singapore), National University of Singapore,
Kent Ridge, 119260, Singapore}
\address{$^3$ System Science Institute, School of Traffic and Transportation,
 MOE Key Laboratory for Urban Transportation Complex Systems Theory
and Technology, State Key Laboratory of Rail Traffic Control and
Safety, Beijing Jiaotong University, Beijing, 100044, P. R. China}
\address{$^4$
Department of Physics \& Astronomy, University of British Columbia,
Vancouver, B.C. Canada, V6T 1Z1} \ead{jinshanw@phas.ubc.ca}

\begin{abstract}
Global degree/strength based preferential attachment is widely used
as an evolution mechanism of networks. But it is hard to believe
that any individual can get global information and shape the network
architecture based on it. In this paper, it is found that the global
preferential attachment emerges from the local interaction models,
including distance-dependent preferential attachment (DDPA) evolving
model of weighted networks(\emph{M. Li et al, New Journal of Physics
8 (2006) 72}), acquaintance network model(\emph{J. Davidsen et al,
Phys. Rev. Lett. 88 (2002) 128701}) and connecting
nearest-neighbor(CNN) model(\emph{A. V\'{a}zquez,  Phys. Rev. E 67
(2003) 056104}).  For DDPA model and CNN model, the attachment rate
depends linearly on the degree or vertex strength, while for
acquaintance network model, the dependence follows a sublinear power
law.  It implies that for the evolution of social networks, local
contact could be more fundamental than the presumed global
preferential attachment.
\end{abstract}

%Uncomment for PACS numbers title message
% Keywords required only for MST, PB, PMB, PM, JOA, JOB?
%\vspace{2pc}
%\noindent{\it Keywords}: Article preparation, IOP journals
% Uncomment for Submitted to journal title message
%\submitto{\JPA}
% Comment out if separate title page not required
\maketitle

\section{Introduction}
\label{Introduction}

Power law distribution commonly exists in complex networks,
including planned network such as Internet\cite{internet} and
unplanned social networks, such as scientific collaboration
networks\cite{Newmanc,Physica,Barrat,Fan}, actor collaboration
networks\cite{actor}, peer-to-peer networks\cite{F.Wu}, mobile
networks\cite{CN:Mobile} and much more. Their degree distributions
obey power law, and if applicable, vertex strength(summing the
weights of links that connect to a vertex)\cite{Barrat,Fan} and link
weight\cite{Collaboration} also follow power law. Many models have
been proposed to capture the topological evolution of complex
networks (see reviews \cite{review,report}). Especially a class of
models based on the idea of global preferential attachment, firstly
proposed in BA model\cite{BAmodel}, is quite successful to reproduce
the power law distributions of
degree/strength\cite{Collaboration,BAmodel,weighted,Traffic}. From
the degree-based preferential attachment in BA model, later
Barrat\cite{weighted} developed a model of weighted networks with
strength-based preferential attachment.

Some networks have hubs or data centers, where global information is
collected for future use. For example, WWW has search engines and
Internet has routers and DNS servers. For networks with such
centers, global preferential attachment mechanism may seem
reasonable. However, in social network there is no data centers
collecting and providing global information. It is impossible for
every individual to know global information of system, such as
degree or strength of every individual. For social networks,
evolution mechanism based on local quantities would be more natural.

Actually, there are already several evolving models for scale-free
networks based on local rules, such as distance-dependent
preferential attachment(DDPA) model\cite{model}, acquaintance
network model\cite{acquaintance}, connecting nearest-neighbor(CNN)
model\cite{localrule}, random walk model\cite{randomwalkers,self},
redirection model\cite{KR,localstrategies}, optimization
model\cite{Exchanges,TPA} and so on.  For instance, in DDPA model a
measure of closeness relation is defined locally within second
neighbors and then links are built up preferentially according to
the relation. Acquaintance network model evolves via people
introduced to know each other by a common
acquaintance\cite{acquaintance}. In CNN model\cite{localrule}, at
every time step, a new vertex is added or a potential edge within
second neighbors is converted into an edge.  All above models based
on local rules can also reproduce the common topological characters
of complex networks.

We notice that both the two potentially conflicting types of models
successfully capture the main characters of scale-free networks, but
one needs global information while the other only requires local
information. Here we suggest a wild guess to resolve the conflict
and make the whole picture more consistent. We conjecture that
global preferential attachment emerges from local contact based
models. In this paper, we are going to investigate whether or not
the above conjecture holds in the following local models: DDPA
model\cite{model}, acquaintance networks model\cite{acquaintance}
and CNN model\cite{localrule}.

This paper is organized as follows. In section \ref{method}, we
first briefly describe the method used to check preferential
attachment. In section \ref{model}, we apply this method to DDPA
model, acquaintance networks model and CNN model, and present the
results. At last, in Section \ref{conclud} we give some concluding
remarks. And it turns out our conjecture does hold in the above
models.

\section{Methods for Measuring Preferential Attachment}\label{method}
The preferential attachment hypothesis states that the rate $\Pi(k)$
with which a vertex with $k$ links acquires new links is a
monotonically increasing function of $k$\cite{BAmodel}, namely
\begin{equation}
\Pi(k_i)=\frac{k_i^ \alpha}{\sum_{j}k_j^ \alpha}=C(t)k_i^
\alpha.\label{preferential}
\end{equation}
For BA model $\alpha=1$\cite{BAmodel}.

H. Jeong et al\cite{measure} proposed a method to check global
preferential attachment from data on evolution process. Here we will
briefly describe the method. In evolving process, we record the order of
each node and link joining the system within a relatively short time frame after
the network evolves steadily in a long time. At a large enough time $T_0$, consider
all vertices existing in the system, called $T_0$ vertices. Next
select a time $T_1$ ($T_1>T_0$ and
$\left(T_{1}-T_{0}\right)\ll T_{0}$), add a new vertex at every time
step between $[T_0, T_1]$. Firstly, count the number $N(k)$ of such
vertices with exactly $k$ degree in $T_0$ vertex. Secondly, record
all the vertices in $T_0$ vertex to whom the new links are attached
as $\Omega$. Lastly, count the number of vertices with exactly $k$
degree in $\Omega$ as $A(k)$. At this condition$, \Pi(k, T_0,
T_1)$ will be independent of $T_0$ and $T_1$ but depend on $k$
only\cite{measure}.

A convenient definition of  $\Pi(k, T_0, T_1)$ function could be
defined as,
\begin{equation}
\Pi(k, T_0,
T_1)=\frac{\frac{A(k)}{N(k)}}{\sum_{k'}\frac{A(k')}{N(k')}}.\label{limh}
\end{equation}
According to mean field theory and Eq.(\ref{preferential}), $A(k)=M
\sum_{i}(\delta(k_{i},k)k_i^{\alpha}/\sum_j k_j^{\alpha})$, where
$M$ is the number of new links attaching to vertices $\Omega$. So
Eq.(\ref{limh}) can be written as
\begin{equation}
\Pi(k)=\frac{M\sum_{i}\delta(k_{i},k)\frac{k_i^{\alpha}}{\sum_l
k_l^{\alpha}}/N(k)}{\sum_{k'}(M\sum_{j}\delta(k_{j},k^{'})\frac{k_j^{\alpha}}{\sum_mk_m^{\alpha}}/N(k'))}
=
\frac{\frac{k^{\alpha}}{\sum_{l}k_{l}^{\alpha}}}{\sum_{k'}\frac{k'^{\alpha}}{\sum_{m}k_{m}^{\alpha}}}
=\frac{k^{\alpha}}{\sum_{k'} k'^{\alpha}}, \label{master}
\end{equation}
where $\sum_{i}\delta(k_{i},k)k_i^{\alpha} = k^{\alpha}N(k)$ is
used. To avoid the effect of noise, we also study the cumulative
function instead of $\Pi(k)$
\begin{equation}
\kappa(k)=\int^{k}_{0} \Pi(k')dk'.
\end{equation}
If there is a global preferential attachment, then
\begin{equation}
\kappa(k)\propto k^{\alpha + 1}.
\end{equation}

\begin{figure}
\center \includegraphics[width=0.45\linewidth]{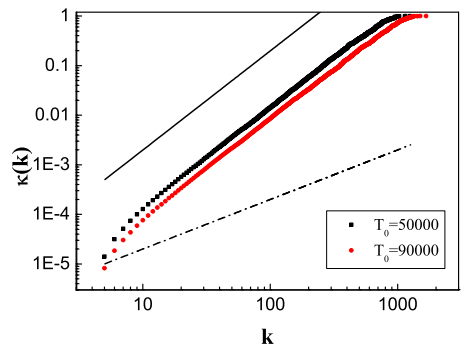}
\caption{The $\kappa(k)$ function determined numerically for BA
model based on Equation (\ref{limh}). The slope of solid line is
$2$(corresponding to $\alpha=1$ as $\alpha+1=2$), and the slope of
dash line is $1$ ($\alpha=0$). This indicates that the exponent
$\alpha$ is almost $1$. If not mentioned in the following figures,
we run the model for $100$ times with the same parameters to count
$A(k,T_0)$ and $N(k,T_0)$, and then we determine
$\kappa(k)$($\kappa(s)$) according to Eq. (\ref{limh}) and estimate
the value of $\alpha$, where the power-law exponents are calculated
on intermediate degree (vertex strength). We used $T_1$ = $T_0+100$
if not mentioned. The parameters are $n_0=10$ and $m=5$.}
\label{measureBA}
\end{figure}

In order to prove the validity of formula (\ref{limh}), we apply
this method on BA model, starting from a fully connected $n_0$
initial network, where $\langle k\rangle=2m$, $\alpha =1$ and
$\gamma =3$. The exponent in Eq.(\ref{master}) should be $1$ and the
measurement shows that they are approximately $1$ (as shown in Fig.
\ref{measureBA}). This indicates that the formula (\ref{limh}) is
fit to measure the form of $\Pi(k)$.

In measurement, we only focus on the vertices new links are attached
onto, so we only consider ending vertices. Furthermore, measurements
on external links from the new vertex and internal links
among existing vertices, can be done separately when necessary.

\section{Measuring preferential attachment of Local Contact Model}

\label{model}
In this section, we are going to check the possibility of such
global attachment emerging from the following three local contact
models.

\emph{\textbf{DDPA model}}: The evolution of networks starts from a
fully connected $n_0$ initial network. At every time step, one new
vertex is added into the network by connecting it randomly to one
old vertex. And then other $l$ old vertices are randomly activated.
Every one (denoted as vertex $n$) of these $1+l$ vertices can
attempt to build up $m$ connections. The probability for every link
(except the first link from the new vertex) from a vertex $n$ to a
vertex $i$ is given by
\begin{equation}
\Pi_{n\rightarrow i} =
\frac{L_{ni}}{\sum_{j\in\partial^{2}_{n}}L_{nj}}, \label{prob}
\end{equation}
where $\partial^{2}_{n}$ means neighbors of vertex $n$ up to the
second separation. $L_{ni}$ is the similarity distance\cite{Fan,
Collaboration}, which means that the larger is the similarity
distance, the closer is the relation between the two end vertices.
If two vertices $i$ and $j$ are connected directly, $L_{ij}=w_{ij}$,
where $w_{ij}$ is the similarity weight, meaning the larger the
closer, such as the number of collaboration in scientific
collaboration networks\cite{Newmanc,Physica,Barrat,Fan}, and the
duration of calls in mobile networks\cite{CN:Mobile}. Otherwise,
$L^{(\mu)}_{ij}$ is calculated as follows. If vertex $i$ ($\mu$) is
connected to vertex $\mu$ ($j$) with similarity link weight
$w_{i\mu}$ ($w_{\mu j}$) then $L^{(\mu)}_{ij}=
\frac{1}{\frac{1}{w_{i\mu}}+\frac{1}{w_{\mu j}}}$. The final
similarity distance between vertex $i$ and vertex $j$ is
$L_{ij}=\max\left \{L^{(\mu)}_{ij}\right\}$, the maximum one via all
vertices $\left\{\mu \right\}$. In other words, it is sufficient to
know the degree of familiarity between second neighbors. In
addition, the reconnection between linked vertices is allowed. When
building up connection between vertices $i, j$ represents an event
happening between the vertices, the number of occurrences of such an
event may be defined as the connection count $T_{ij}$, which will be
transferred into similarity link weight $w_{ij}=f(T_{ij})$, e.g.,
$w_{ij}=T_{ij}$. This DDPA mechanism does not use vertex degree or
vertex strength, i.e., $s_i=\sum_j w_{ij}$, as the reference, not
even locally. It only makes use of the information about local
similarity distance within second neighbors, but scale-free
phenomenon does appear in DDPA model for intermediate
degrees/strengths(see Fig.6 in Ref.\cite{model}).

\begin{figure}
\center
\includegraphics[width=0.45\linewidth]{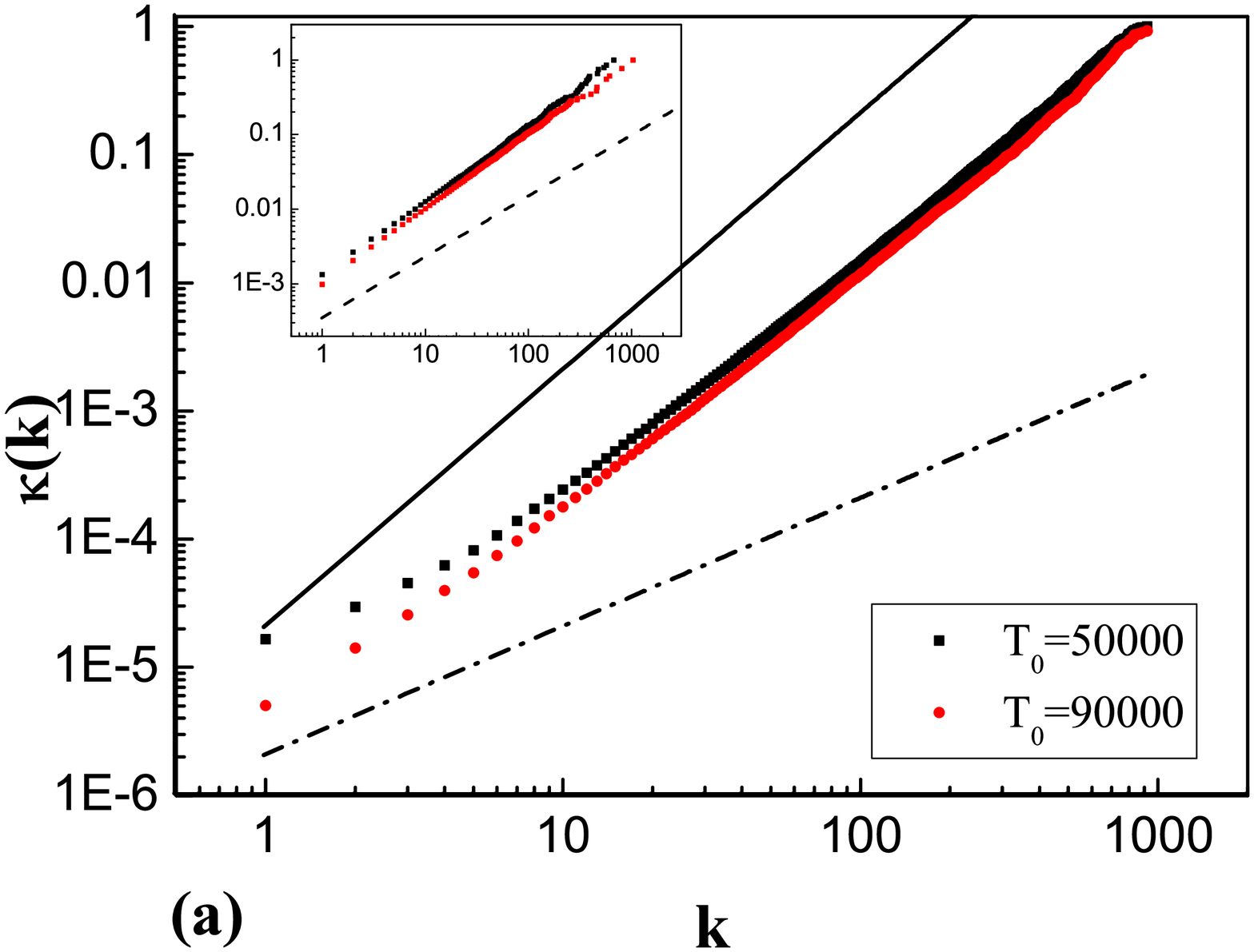}
\includegraphics[width=0.45\linewidth]{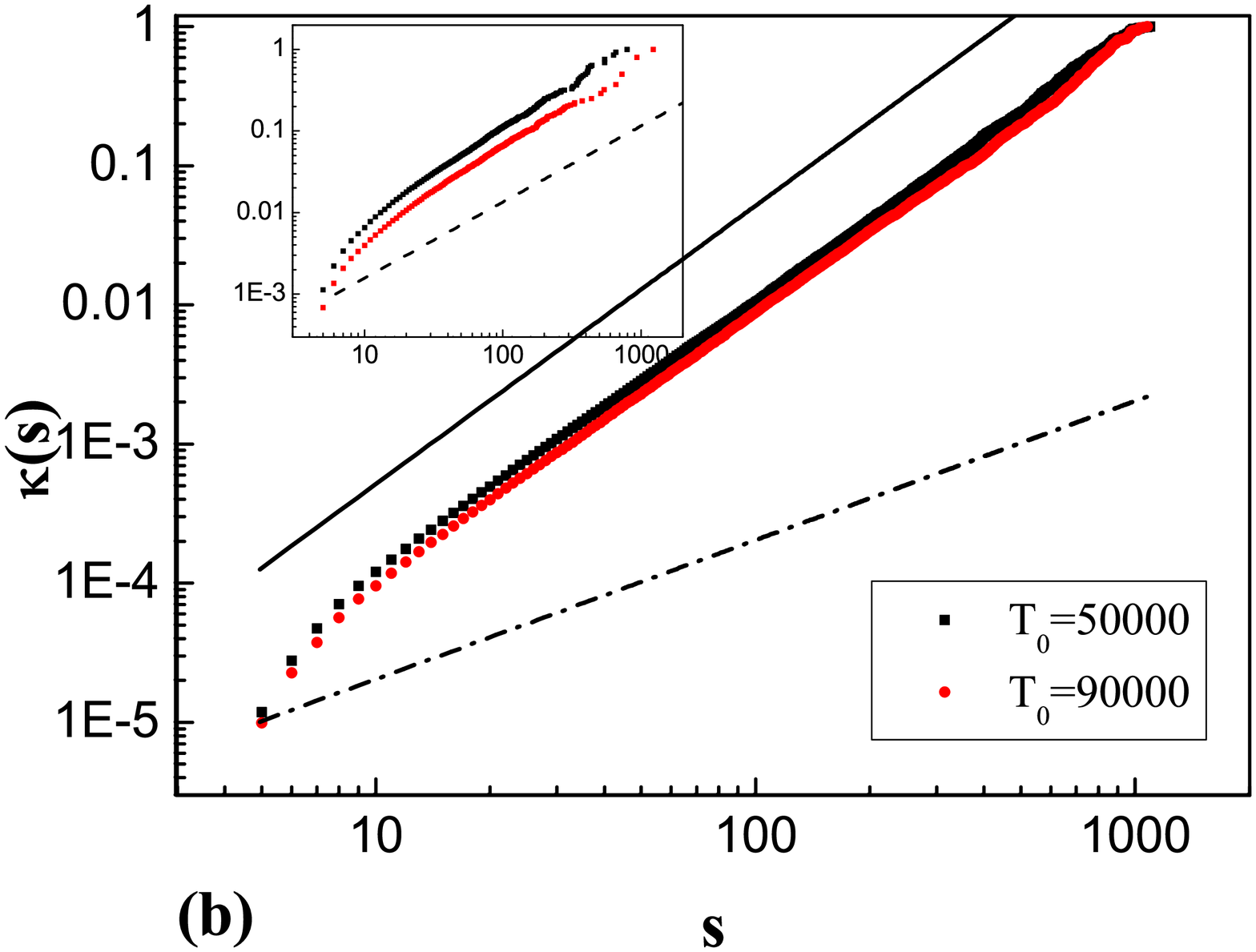}
\caption{The $\kappa(k)$ (a) and $\kappa(s)$ (b) function determined
numerically for external links of DDPA model.  The insets are
results for the first external links, where $\alpha_{k}=0.01$ and
$\alpha_{s}=0.05$, almost flat distributions. And $\langle \alpha_k
\rangle= 0.85$(a) and $\langle \alpha_s \rangle= 0.91$(b) for all
other external links. The parameters of DDPA model are $n_0=10$(the
size of initial networks), $l=1$(the number of activated old vertex)
and $m=5$(the number of new connections on every activated vertex).}
\label{measureex}
\end{figure}

For DDPA model, we need to do this measure for three kind of links:
the first external links from the new vertex, all other external
links from the new vertex and internal links. The same method can
also be applied to check preferential attachment according to vertex
strength. Since the ending vertex of first external link is selected
randomly, the probability of a vertex selected should be independent
of its degree ($k$) or strength($s$).  In the inset of
Fig.\ref{measureex}, we see the first external link does follow flat
distribution. But for all other external links, we find that
$\kappa(k)$ (measuring according to degree) and $\kappa(s)$
(measuring according to strength) for intermediate degrees/strengths
follow a straight line on a log-log plot, indicating $\Pi (k)\propto
k^{\alpha_{k}}$ or $\Pi (s)\propto s^{\alpha_{s}}$. The curves of
$\kappa(k)$ for DDPA model are consistent with those for BA model
(Fig.\ref{measureBA}). And they are parallel with different $T_0$
(as shown in Fig.\ref{measureex}), indicating that $\Pi(k)$ and
$\Pi(s)$ are independent of $T_0$ and $T_1$, and depend on $k$ or
$s$ only. Their power law exponents are respectively $\langle
\alpha_k \rangle=0.85$ and $\langle \alpha_s \rangle= 0.91$.

\begin{figure}
\center
\includegraphics[width=0.45\linewidth]{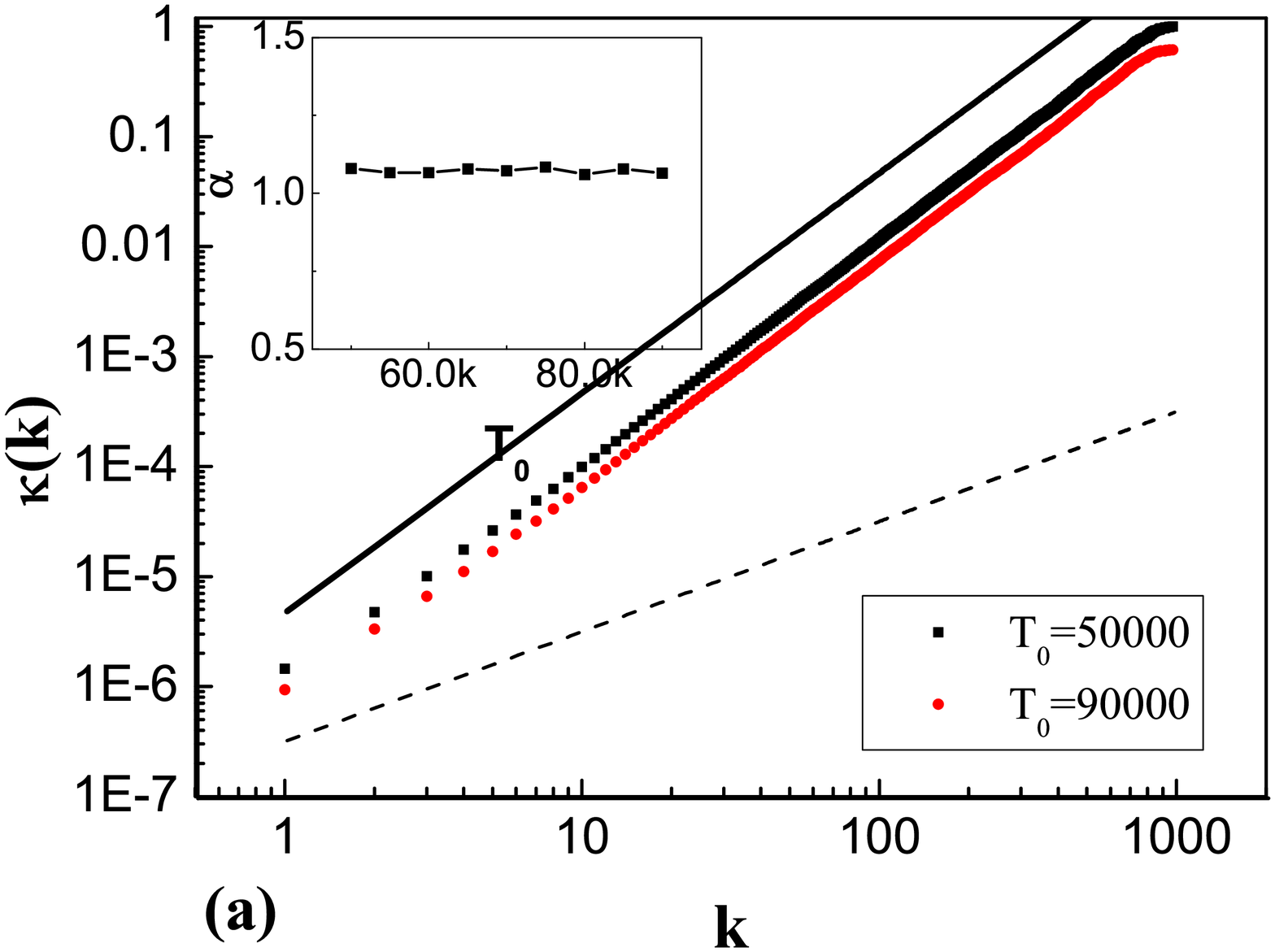}
\includegraphics[width=0.45\linewidth]{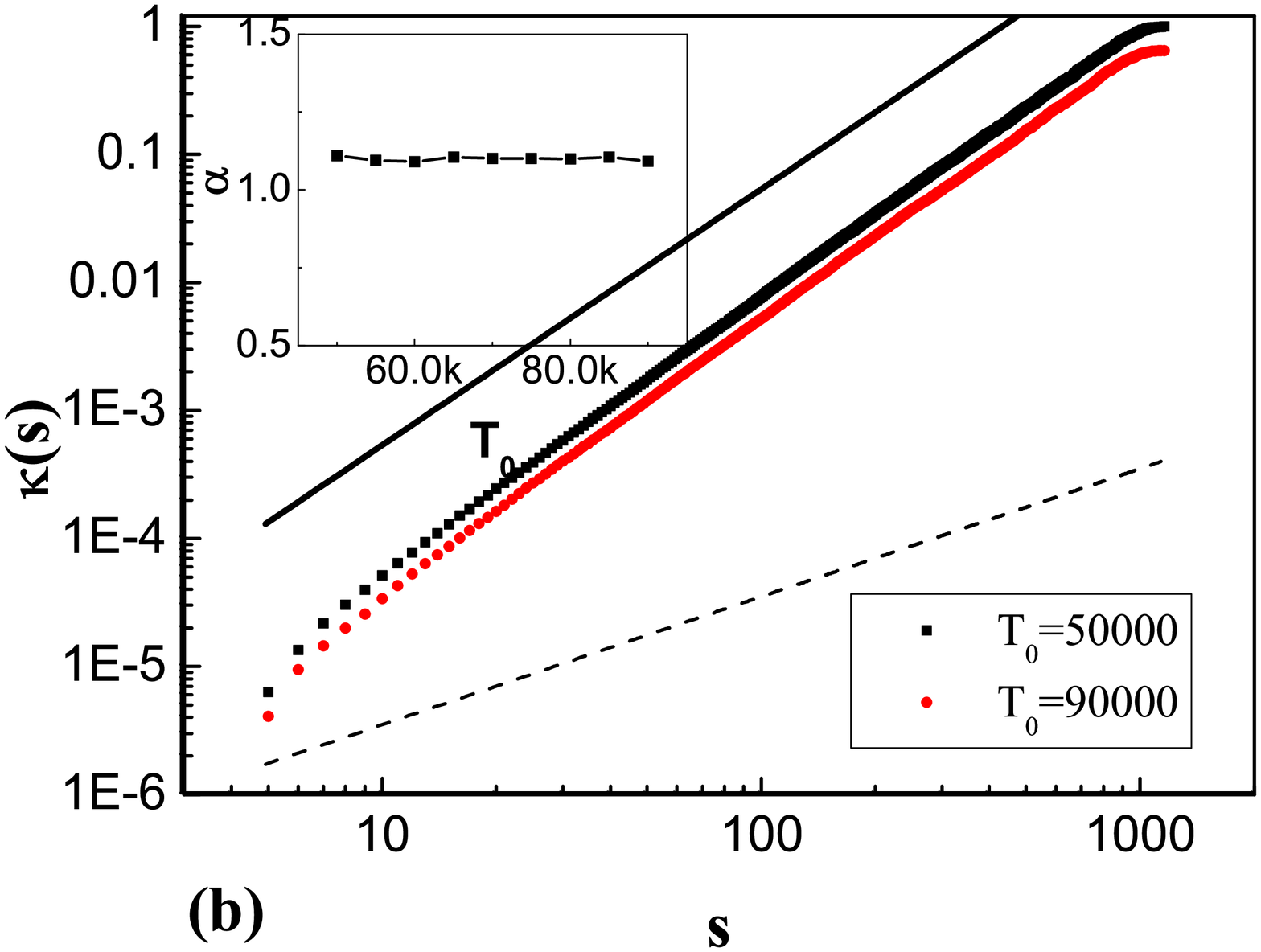}
\caption{The $\kappa(k)$ (a) and $\kappa(s)$ (b) function determined
numerically for internal links of DDPA model. The curves increase
faster than linear(shown as dashed line in the figures), showing
that the probability of vertex linked by internal link is
proportional to degree and vertex strength with $\langle \alpha_k
\rangle= \textbf{1.07}$(a) and $\langle \alpha_s \rangle=
\textbf{1.1}$(b). Inserted are the values of $\langle \alpha
\rangle$ for different choice of $T_{0}$. It shows very small
variation.} \label{measurein}
\end{figure}

In the measurement of internal links, we find that global
preferential attachment is also valid. The curves of $\kappa(k)$ and
$\kappa(s)$ follow a straight line on a log-log plot. Their power
law exponents are respectively $\langle \alpha_k \rangle= 1.07$ and
$\langle \alpha_s \rangle= 1.1$ (as shown in Fig.\ref{measurein}).
Actually, if checking all links together, including the first
external link, the curves of $\kappa(k)$ and $\kappa(s)$ follow a
straight line on a log-log plot, where exponents are respectively
$\langle \alpha_k \rangle= 0.98$ and $\langle \alpha_s \rangle=
1.04$ when $l=1$. The parameters almost have no effect on the
exponents $\langle \alpha_k \rangle$ and $\langle \alpha_s \rangle$.

\begin{figure}
\center \includegraphics[width=0.45\linewidth]{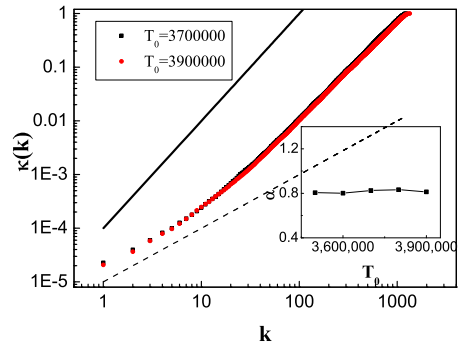}
\caption{The $\kappa(k)$ function determined numerically for
acquaintance network model. Inserted are the values of $\langle
\alpha \rangle$ for different choice of $T_{0}$. Results from
fitting of $\alpha$ shows that $\langle \alpha \rangle=
\textbf{0.815}$. The parameters are $N=10000$, $p_l=0.01$ and
$p=0.01$.} \label{measureacq}
\end{figure}

\emph{\textbf{Acquaintance model}}:  At first, a random network with
$N$ vertices, where every pair vertices are connected with
probability $p_l$, is generated. Then acquaintance networks evolve
according to the following rules: (i) One randomly chosen person
introduces two random acquaintances of his to each another. If they
have not met before, a new link between them is formed. If he has
less than two acquaintances, then he introduces himself to another
random person. (ii) With probability $p$ one randomly chosen person
is removed from the network, including all links connected to this
vertex, and replaced by a new person with one randomly chosen
acquaintance\cite{acquaintance}. The model generates degree
distributions spanning scale-free and exponential regimes(see Fig.1
in Ref. \cite{acquaintance}). Though the size of network is fixed,
this model includes birth and death process. New links emerge
between individuals frequently. After evolving for $T_0$ times,
again we record and analyze evolution process between $T_0$ and
$T_1$. We can see that $\kappa(k)$ also follows a straight line on a
log-log plot, where $\alpha \simeq 0.80$(as shown in Fig.
\ref{measureacq}).

\emph{\textbf{CNN model}}: Next we consider a variant of the
acquaintance model, connecting nearest-neighbor
model\cite{localrule} with increasing network size. CNN model starts
from a single vertex and evolves according to the following rules:
(i) With probability $1-u$, a new vertex(denoted as $n$) is
introduced into the network by connecting it randomly to one old
vertex $j$. The pair of vertices $[n,i]$ between all nearest
neighbors $i$ of vertex $j$ and the new vertex $n$ is recorded in
$\bar{\partial}([n,i])$. (ii) Then, with probability $u$ two
vertices $[n,i]$ selected randomly from $\bar{\partial}$ are
connected. When the size of network is sufficiently large, the
distribution of intermediate degrees exhibits a power-law decay
$P(k) \sim k^{-\gamma}$(see Fig. 8 in Ref. \cite{localrule}), and
the value of $\gamma$ depends on the parameter $u$.

\begin{figure}
\center
\includegraphics[width=0.45\linewidth]{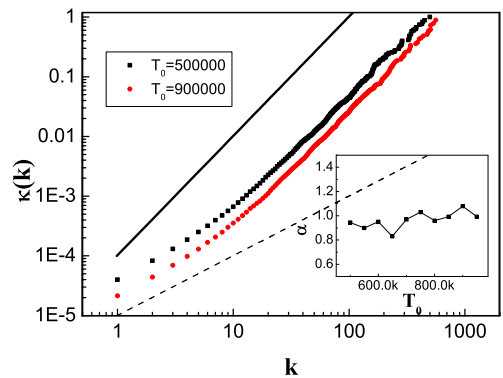}
\caption{The $\kappa(k)$ function determined numerically for all
links of CNN model. Inserted are the values of $\langle \alpha
\rangle$ for different choice of $T_{0}$. For $u=0.4$ we find that
$\langle \alpha \rangle \approx 1.00$.} \label{fig5}
\end{figure}

In the process of measurement, we record the degree of vertex $j$
and the other two selected end vertices. We see that $\kappa(k)$
follows a straight line on a log-log plot, and they are parallel
with different $T_0$(as shown in Fig.\ref{fig5}). For different
parameter $u$, the value of exponent $\alpha$ are different, for
example, $\alpha \approx 1.11$ for $u=0.3$ and $\alpha \approx 0.95$
for $u=0.7$. In Fig.\ref{fig5}, we show the measurement on $u=0.4$.
And there we can see that $\alpha \approx 1.00$.

Although $\alpha$ is close to $1$ for DDPA and CNN model and
$\alpha=0.80$ for acquaintance model, we find that $\kappa(k)\sim
k^{\alpha+1}$ holds in all models. And this indicates that although
links are created by local information, it seems as if the network
evolves according to global preferential attachment.

\section{Concluding Remarks}\label{conclud}
In this paper, we measured global degree/strength preferential
attachment of DDPA model, acquaintance network model and CNN model.
These models make use of local quantities instead of global
degree/strength preferential attachment mechanism. However, our
measurement shows that they still can be seen as if the networks
evolving according to global preferential attachment. From this
point of view, preferential attachment is an emergent phenomenon
from the more fundamental local rules. Empirical study on e-mail
networks in Ref.\cite{Science_311_88} also suggests that cyclic
closure, especially triadic closure, plays an important role in
building up new links. In modeling social networks, this observation
may suggest that local rules are preferred rather than the global
degree/strength preferential attachment. For people who believe in
the later, our discovery is especially meaningful. Our results
indicate that one does not need to worry about the widely believed
and used global preferential attachment because it can emerge from
properly designed local rules.

\section*{Acknowledgments}
The authors are very grateful to the anonymous referees for helpful
comments and suggestions. This work is partially supported by $985$
Project, NSFC under the grant No. $70771011$ and No. $60974084$. M.
H. Li is also supported by DSTA of Singapore under Project Agreement
POD0613356. L. Gao thanks for National Basic Research Program of
China (No. 2006CB705500) and Project of National Natural Science
Foundation of China (No. 70631001).

\end{document}